# Spin dephasing in III-V nanowires

Ashish Kumar[a)] and Bahniman Ghosh

*Department of Electrical Engineering, Indian Institute of Technology, Kanpur 208016, India*

Abstract – We use semiclassical Monte Carlo approach to investigate spin polarized transport in InP and InSb nanowires. Spin dephasing in III-V channels is caused due to D'yakonov-Perel (DP) relaxation and due to Elliott-Yafet (EY) relaxation. The DP relaxation occurs because of bulk inversion asymmetry (Dresselhaus spin-orbit interaction) and structural inversion asymmetry (Rashba spin-orbit interaction). The injection polarization direction studied is that along the length of the channel. The dephasing rate is found to be very strong for InSb as compared to InP which has larger spin dephasing lengths. The ensemble averaged spin components vary differently for both InP and InSb nanowires. The steady state spin distribution also shows a difference between the two III-V nanowires.

Keywords: Spin Dephasing, Monte Carlo, InP, InSb, Nanowires,

a). Electronic mail : ashish12.kumar@gmail.com

1. INTRODUCTION

The multidisciplinary field of spintronics [1-4] has attracted significant interest from the researchers over the years. This has resulted in a gradual increase in exploring the spin degree of freedom [2-6] as opposed to the charge degree of freedom. Spin transport in semiconductors has gained momentum due to the prospects of integrating spintronics with conventional electronics to implement novel devices [7-11] . These novel devices have the capability of operating at higher data processing speeds with decreased power dissipation and at lower power levels. A number of such devices have also been proposed. Moreover semiconductor based spintronics can help integrate storage, logic, communication on a single chip thereby providing a multifunctional device [1]. Also spin, being a quantum operator, can be used for quantum computation [12-14].

There are three fundamental aspects governing the operation of a spin based device – how to effectively polarize a spin system, how long is the system able to retain its system polarization, how can spin be efficiently detected. These three aspects lead us to the three processes that form the core of the study on spintronics-spin injection, spin relaxation and spin detection. Research has shown that spin orientation of electrons in semiconductors is preserved for a much longer time [15, 16] than momentum. Apart from spin dephasing time, another spin property, spin dephasing length becomes a determining property when transmitting information needs to be conserved. Our work here focuses the second process of spin dephasing.

In this paper we study spin polarized transport in III-V nanowires. Of late, III-V compounds are a centre of intensive research. This is attributed to the fact that III-V compounds possess properties that are more suited for ideal device charactereistics. They have high carrier mobility and high saturation velocity. This helps in manufacturing high frequency devices. Also being direct semiconductors, they are suitable for optical applications. Thus such materials can be used to transmit optically coded information via electron spin. One of the biggest advantages that III-V devices provide is the ability to alter the bandgap to maximize performance for a particular operation. Being such a potent semiconductor material, spin transport in such materials has been an area of huge research. A lot of research, both experimental and theoretical has already been done and a lot is still being pursued to comprehend their spin transport properties. In Ref. [17] GaAs is studied experimentally to ascertain its spin transport properties using a spectroscopic method. In Ref. [18], spin polarized transport in III-V quantum well is studied using a single subband approximation. In Ref. [19] Monte Carlo method is used to simulate spin polarized transport in GaAs/GaAlAs quantum wells. Spin dephasing is studied at different electric field, temperatures and channel widths. In Ref. [20] a multisubband Monte Carlo method is used to determine spin dephasing lengths in GaAs nanowires In Ref. [21] spin dephasing times are determined in GaAs quantum wires using multisubband Monte Carlo approach. In Ref. [22] spin polarized transport in 1D and 2D $In_{0.53}Ga_{0.47}As$ heterostructures is compared. In Ref. [23] spin transport in InSb/InAlSb 2D heterostructures is investigated using a single subband approach. In a bid to examine materials [24] suited for spintronics applications, we undertake our study on InP and InSb nanowires.

In this work, we simulate spin polarized transport in InP and InSb nanowires. A number of classical drift diffusion models [25, 26] and fully quantum mechanical [11, 27] models have already been developed to study spin transport. However, they suffer with certain inadequacies [21, 28]. Here in our work, we use a multisubband semiclassical Monte Carlo approach to model spin dephasing. Monte Carlo simulations have been widely adopted to study electron transport in devices and have recently been used in conjunction with spin density matrix calculations to model spin transport [18-24]. Monte Carlo approach is best suited for studies on spin dephasing since the spin evolution occurs continuously in step with the evolution of momentum which is taken care of easily by a Monte Carlo simulation.

The paper is organized as follows. The next section deals with the theory and a description of our model. In Section 3 we present the results of our simulations along with the discussion on the results. Finally we conclude in Section 4.

2. MODEL

A full account of the Monte Carlo simulations [18, 29, 30] and spin transport model [18, 20, 21, 23] is described elsewhere. In this paper we shall restrict ourselves to discussing only the necessary features of the model and the key modifications.

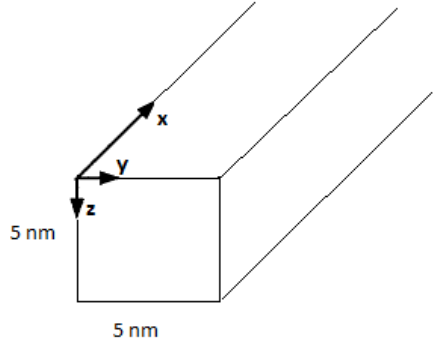

Fig.1 Geometry of the nanowire and the co-ordinate axes.

Fig.1 shows the geometry nanowire structure and the designation of the axes in accordance with the co-ordinate system chosen. The current flow is maintained by the application of a driving electric field $E_x$ along the channel. Along with $E_x$, a transverse field is also present. III-V compounds possess bulk inversion asymmetry which leads to Dresselhaus spin-orbit interaction [31]. The transverse field breaks the structural inversion symmetry which causes Rashba spin-orbit coupling [32]. The electron spin and electron momentum are coupled via spin-orbit interaction. The electron spin evolves under the influence of spin-orbit Hamiltonian which comprises of the Dresselhaus interaction [31] expressed as

$$H_D = -\beta(<k_y>^2 - <k_z>^2)k_x\sigma_x, \qquad (1)$$

and of the Rashba interaction written as [33]

$$H_R = -\eta k_x \sigma_y \qquad (2)$$

The constants $\beta$ and $\eta$ depend on the material. $\eta$ also depends on the external transverse electric field and this dependence is explicit from the expression of $\eta$ [34],

$$\eta = \frac{\hbar^2}{2m^*}\frac{\Delta}{E_g}\frac{2E_g+\Delta}{(E_g+\Delta)(3E_g+2\Delta)}eE \qquad (3)$$

where $\Delta$ is the spin orbit splitting of the valence band, $e$ is the electronic charge, $m^*$ is the effective mass, $E_g$ is the band gap and $E$ is the transverse electric field.

The temporal evolution of the spin vector during the free flight time occurs in accordance with the following equation [20, 21],

$$\frac{d\vec{S}}{dt} = \vec{\Omega} \times \vec{S}. \tag{4}$$

The so-called "precession vector" $\vec{\Omega}$ has contributions from the Dresselhaus interaction and f the Rashba interaction and can be written as [20,21],

$$\Omega_D(k_x) = -\frac{2\beta_{eff} k_x \hat{\imath}}{\hbar} \tag{5}$$

$$\Omega_R(k_x) = -\frac{2\eta k_x \hat{\jmath}}{\hbar} \tag{6}$$

where $\beta_{eff} = \beta(<k_y>^2 - <k_z>^2)$.

Using Eq. (5) and Eq. (6) in Eq. (1) and expressing spin vector $\vec{S} = S_x \hat{\imath} + S_y \hat{\jmath} + S_z \hat{k}$, we get the following relations for the individual components of spin,

$$\frac{dS_x}{dt} = -\frac{2}{\hbar} \eta k_x S_z \tag{7}$$

$$\frac{dS_y}{dt} = \frac{2}{\hbar} \beta_{eff} k_x S_z \tag{8}$$

$$\frac{dS_z}{dt} = \frac{2}{\hbar} k_x (\eta S_x - \beta_{eff} S_y) \tag{9}$$

The entire simulation time is divided into small time steps $\Delta t$ and the spin components are updated after every such time step. During $\Delta t$ the spin dynamics is coherent without any dephasing since the evolution is unitary during this interval. However the presence of driving electric field and scattering events change the electron wavevector and produce a distribution of momentum states. These in turn result in a distribution of spin states in the ensemble leading in ensemble dephasing. This is the D'yakonov-Perel (DP) [35] relaxation. There is yet another type of dephasing mechanism, Elliott-Yafet (EY) [reference for EY] relaxation. EY relaxation [36] causes instantaneous spin flip and is treated as a spin flip scattering. The spin relaxation time is given by [37],

$$\frac{1}{\tau_s^{EY}} = A \left(\frac{k_B T}{E_g}\right)^2 \alpha^2 \left(\frac{1-\alpha/2}{1-\alpha/3}\right)^2 \frac{1}{\tau_p} \tag{10}$$

where $E_g$ is the band gap, $\alpha = \Delta/(E_g + \Delta)$ where $\Delta$ is the spin orbit splitting and $\tau_p$ is the momentum relaxation time. A is a dimensionless constant and varies between 2 and 6. For this work we have chosen A as 4.

The conduction band [38] in III-V compounds is characterized by a Γ-valley minimum along with L-valley and X-valley which are higher up in energy than the Γ-valley. For the sake of our simulation we only take the lowermost Γ-valley into account and assume the other two valleys to be depopulated being higher up in energy levels.

The scattering mechanisms taken into account in our simulations are acoustic phonon scattering, surface roughness scattering, ionized impurity scattering and polar optical phonon scattering. The formulae for computation of scattering rates are taken from references [39, 40, 41].

## 3. RESULTS AND DISCUSSION

We have used the model described in the preceding section to simulate spin polarized electron transport in InP and InSb nanowires. The nanowire structure considered is of cross-section 5nm x 5nm. The transverse effective field is taken to be 100 kV/cm. This effective field results in Rashba spin orbit coupling. Four subbands[24] are considered in the simulations to account for the confinement along the two transverse directions. The higher subbands will be very higher up in energy due to small transverse dimensions and hence for the purpose of the simulation they can be considered to be depopulated and thus are neglected. Moreover, the moderate values of driving electric field (1kV/cm) used in our simulation ensure that the majority of electrons are contained in the first four subbands [22,24]. The energy levels of subbands are calculated using an infinite potential well approximation. The Rashba coefficient $\eta$ is calculated using the Eq (3) for both InP and InSb. The spin orbit splitting for InP and InSb is 0.11 eV and 0.80 eV respectively [42]. The Dresselhaus coefficient values for InP and InSb are taken from [42]. The material parameters for Monte Carlo simulation for InP are taken from Ref. [43] and for InSb are taken from Ref. [44]. The electrons are injected with a specific polarization from the source i.e. x=0. A time step of 0.02 fs was selected and the simulation run for 12 x $10^5$ such time steps. This allows the electrons to reach steady state. Data is recorded for the final 50,000 steps only. The ensemble average is calculated for each component of the spin vector for the last 50,000 steps at each point of the wire according to the expression [20],

$$<S_i>(x,T) = \frac{\sum_{t=t_1}^{t=T} \sum_{n=1}^{n_x(x,t)} S_{i,n}(t)}{\sum_{t=t_1}^{t=T} n_x(x,t)} \qquad (11)$$

Here $i$ denotes the $x$, $y$ and $z$ components, $n_x(x,t)$ is the total number of electrons in a grid of distance $\Delta x$ around position $x$ at time $t$, $S_{i,n}(t)$ represents the value of the $i^{th}$ spin component of the $n^{th}$ electron at time $t$. Here $T$ is the end time and $t_1$ is the time at which we start recording the data. The magnitude of the average spin vector is then computed using the expression

$$|<S>(x,T)| = \sqrt{<S_x>^2 + <S_y>^2 + <S_z>^2} \qquad (12)$$

Spin dephasing length is defined as the distance from the source (x=0, from which the electrons are injected) where |<S>| drops to 1/e times of its initial value of injection. In our simulations the electrons are injected with an initial polarization of 1 and hence the initial value of |<S>| is 1.

A. *Decay of magnitude of ensemble averaged spin vector for InP and InSb nanowires*

Figure 2 shows the decay of the magnitude of ensemble averaged spin along a InP nanowire at 300K and driving electric field of 1kV/cm. The electrons are injected with spin polarized along the length of the channel (axis of the nanowire) i.e. along the x-direction. The spin dephasing length is 20.93 µm.

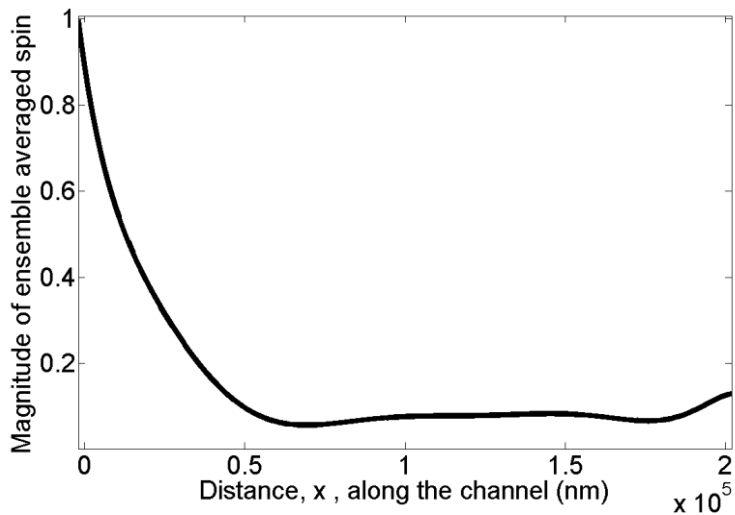

Fig.2. Decay of spin along a InP nanowire for injection polarization along the x-direction at 300K a driving electric field of 1kV/cm

Figure 3 shows the decay of the magnitude of ensemble averaged spin along a InSb nanowire at 300K and driving electric field of 1kV/cm. The electrons are again injected with spin polarized along the length of the channel. The spin dephasing length is 320nm.

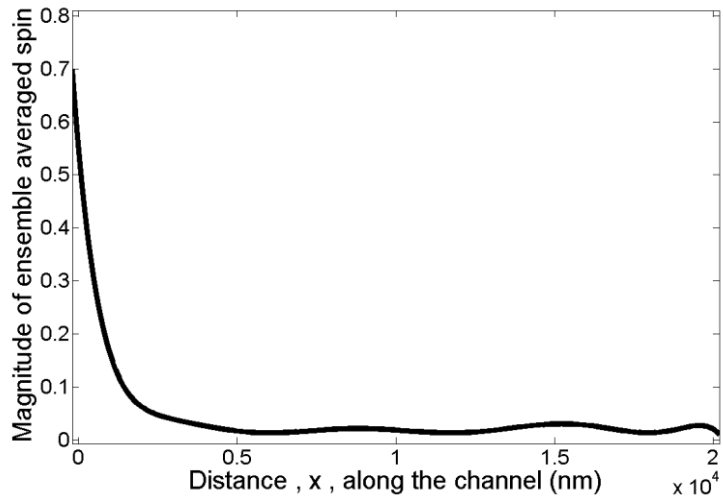

Fig.3. Decay of spin along a InSb nanowire for injection polarization along the x-direction at 300K a driving electric field of 1kV/cm

There is a huge difference in the spin dephasing lengths in InP and InSb. To quantize, the spin dephasing length in InP is approximately 65 times longer than in InSb. This difference is due to the difference in depolarization rates in InP and InSb because of DP relaxation and EY mechanisms. The spin orbit coupling in InSb is much larger than in InP. The value of Rashba coefficient at 300K and at the transverse electric field 100kV/cm from Eq. (3) for InP is $2.78 \times 10^{-32}$ and for InSb is $1.12 \times 10^{-29}$. The value of Dresselhaus spin-orbit parameter $\beta$ [42] used for simulations is $8.5\ eV-\text{Å}^3$ for InP and $220\ eV-\text{Å}^3$ for InSb. Thus both the Rashba and Dresselhaus spin orbit interaction is stronger in InSb resulting in stronger DP relaxation and thus faster dephasing in InSb. Also InSb is a narrow gap semiconductor (0.17 eV) with very high spin orbit coupling (0.80eV) whereas InP is a wide bandgap (1.34 eV) semiconductor with a weak spin orbit coupling (0.11eV). Thus the Elliott Yafet spin relaxation mechanism is strongly dominant in InSb while it is much weaker in InP. This also leads to faster depolarization in InSb. Thus the spin dephasing lengths are longer in InP compared to InSb.

B. *Decay of spin components in InP nanowire*

Figure 4 shows the decay of the ensemble averaged *x*, *y* and *z* components of the spin vector along the InP nanowire for *x*-polarized injection. The driving electric field is 1kV/cm and the temperature is 300K.

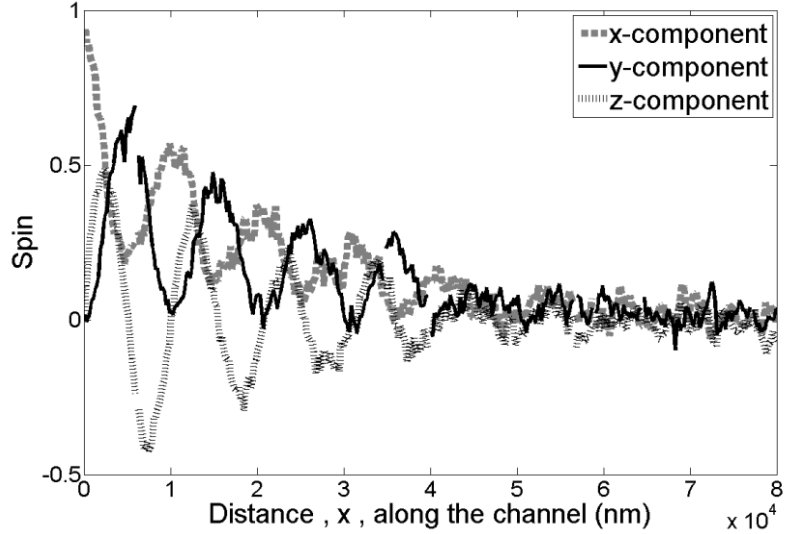

Fig.4. Dephasing of the *x,y* and *z* components of ensemble average spin in InP nanowire at 300K at a driving electric field of 1kV/cm with initial injection polarization along the *x*-direction

In Section 2 we had discussed that spin evolution can be split into coherent dynamics (rotation of spin vector which is unitary) and incoherent dynamics (spin depolarization or decay in magnitude) [21]. There is a constant competition between these two processes determined by the relative strengths of the dephasing rates and the spin precession vector. The coherent motion manifest itself in the form of a oscillatory component in the spatial profile of the spin components while the dominance of dephasing rates leads to a monotonic decay.

From the Figure 4 we note that the spatial decay of the spin components for InP has a oscillatory component for the *x*-polarized injection. Based upon the above reasoning, this leads us to the inference that the incoherent dynamics due to ensemble dephasing is weaker than the coherent dynamics in InP. The coherent dynamics dominates because the spin precession vector is large due to large spin orbit coupling in InP.

Since the coherent dynamics is dominant over the dephasing rates, the Eqs. 7, 8 and 9 can be solved to ascertain analytically the dephasing profiles depicted in Fig.4.

## C. Decay of spin components in InSb nanowire

Figure 4 shows the decay of the ensemble averaged *x*, *y* and *z* components of the spin vector along the InP nanowire for *x*-polarized injection. The driving electric field is 1kV/cm and the temperature is 300K.

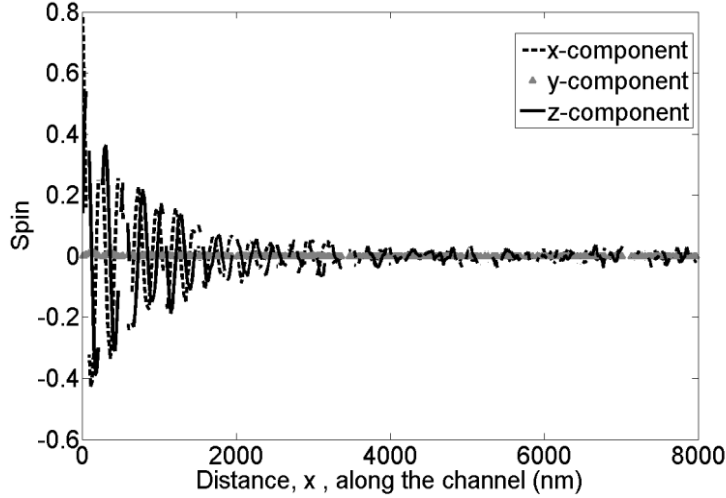

Fig.5. Dephasing of the *x*,*y* and *z* components of ensemble average spin in InSb nanowire at 300K at a driving electric field of 1kV/cm with initial injection polarization along the x-direction.

Since the initial polarization is along the *x*-direction, the ensemble averaged *y*-component of the spin vector remains near zero. This is because the Rashba interaction does not couple the *x*- or *z*- polarized spins to the *y*-polarized spins. Only the Dresselhaus interaction is coupled to the y-polarized spin. Any change in the value of the *y*-component of spin occurs only because of this Dresselhaus interaction. However the Dresselhaus interaction for for InSb nanowire is very weak compared to the Rashba interaction and thus the *y*-component of spin (with an initial value of 0 at x=0) remains near zero. The *x*-component start with the value 1 at x=0 since this is the injected polarization while the *z*-component starts with an initial value of 0. The initial phase difference of $\pi/2$ between the *x*- and *z*- components changes due to dephasing.

Also from Figure 5 we observe that the *x*- and *z*- component of spin display an oscillatory behaviour which is indicative of the fact that for *x*-polarized injection in InSb the coherent dynamics dominates over incoherent dynamics. This is true since InSb has a large Rashba spin orbit coupling (much larger than InP) and hence the spin precession vector due to Rashba interaction is very large. Thus spin rotation dominates over ensemble dephasing.

D. *Steady state spin distribution in InP nanowire*

Figure 6 shows the steady state distribution of the *x, y* and *z* components of the spin vector of electrons in the ensemble for *x*-polarized injection in InP nanowire. The driving electric field is 1kV/cm and the lattice temperature is 300K. Apart from slight variations, the spin distribution shows a more or less uniform distribution which indicates that all values of spin are equally likely for the three spin components [21]. The reason being that in InP nanowire, both the expectation values for the Dresselhaus and Rashba terms are close to each other with the Dresselhaus interaction being only slightly stronger than the Rashba interaction. This causes all the three spin components to be coupled uniformly to each other and hence all the spin components are equally likely.

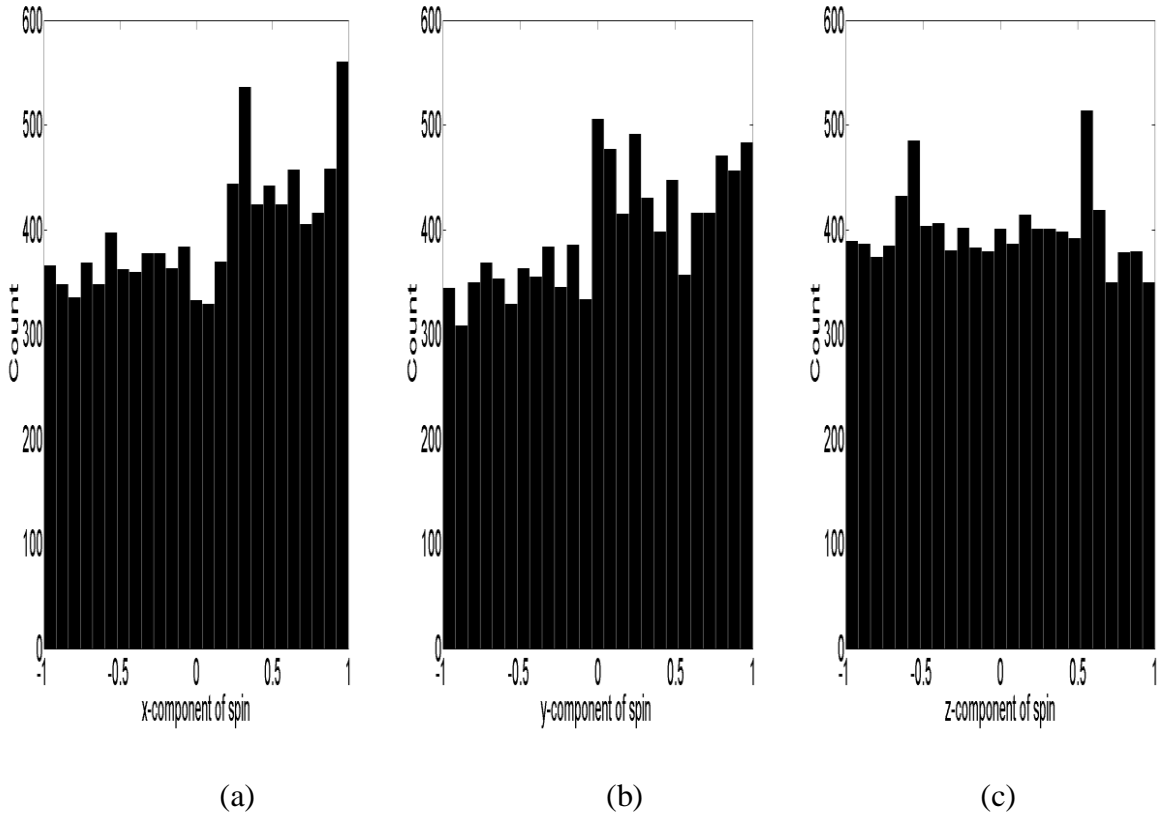

(a)　　　　　　　　　　　(b)　　　　　　　　　　　(c)

Fig.6. Steady state distribution of the spin components in the InP nanowire at 300K at a driving electric field of 1kV/cm with initial injection polarization along the *x*-direction (a) Distribution of the *x*-component, (b) distribution of the *y*-component and (c) distribution of the *z*-component

## E. Steady state spin distribution in InSb nanowire

Figure 7 shows the steady state distribution of the *x, y* and *z* components of the spin vector of electrons in the ensemble for *x*-polarized injection in InSb nanowire. The driving electric field is 1kV/cm and the lattice temperature is 300K.

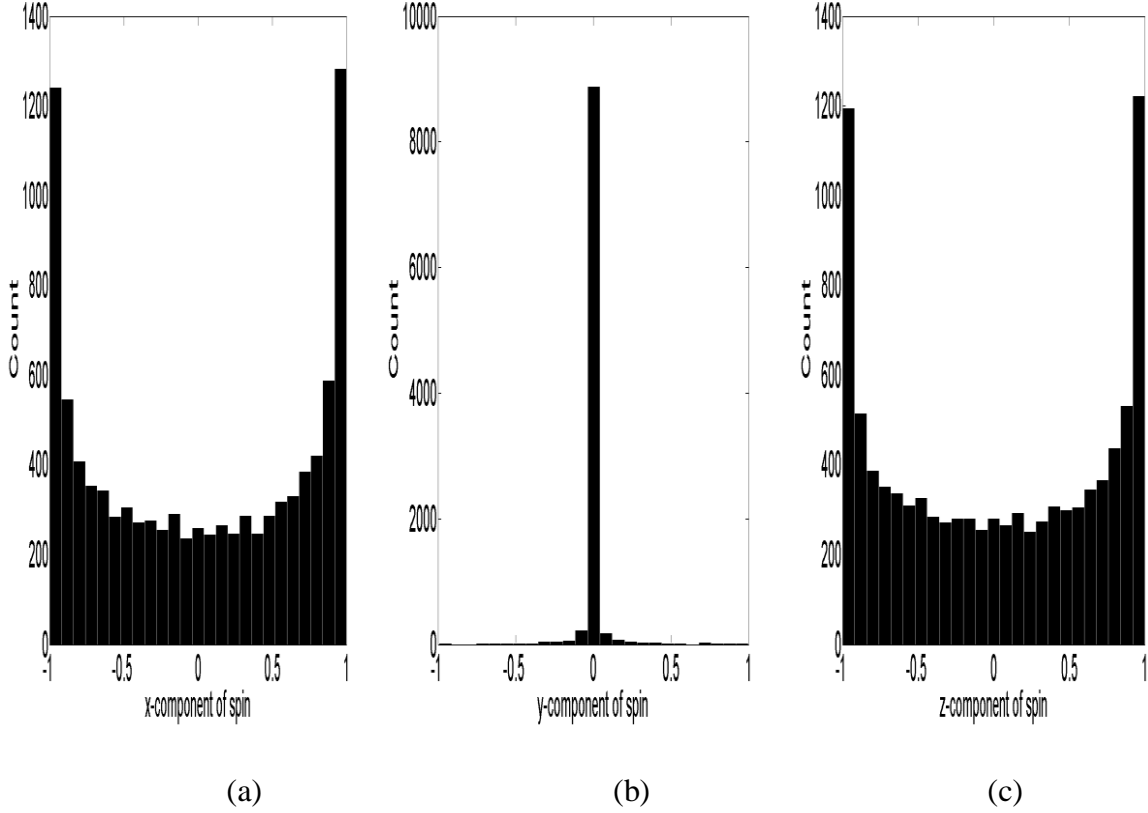

(a)　　　　　　　　　　(b)　　　　　　　　　　(c)

Fig.7. Steady state distribution of the spin components in the InSb nanowire at 300K at a driving electric field of 1kV/cm with initial injection polarization along the *x*-direction (a) Distribution of the *x*-component, (b) distribution of the *y*-component and (c) distribution of the *z*-component

Figure 7(b) shows that the y-component of spin vector is a delta function at zero which means that the y-component remains near zero for x-polarized injection in InSb. This follows directly from our discussion in Section 3(C).

The *x* and *z* components of spin vector show a U-shaped distribution with more electrons in the ensemble having a tendency to have values close to +1 and -1. The U-shape is a manifestation of the oscillatory decay of the *x* and *z* components [21].

4. CONCLUSION

In this work, we studied spin dephasing in a nanowire. The study was conducted on InP and InSb nanowires, being viable III-V materials, in a bid to investigate and compare their spin transport properties. The electrons were injected with initial polarization along the axis of the nanowire, i.e. along the x-direction. The spin dephasing length in InP (20.93 μm) was found to be much longer than in InSb (320 nm) due to rapid spin depolarization in InSb. This suggests that InP nanowire can act as a better one-dimensional channel to transmit information since the spin dephasing rate is slow in InP. The decay of spin components is oscillatory for InP and InSb nanowires. The steady state spin distribution is uniform for InP for all three components. On the other hand, the steady state spin distribution in InSb exhibits a U-shaped profile for *x* and *z* components while for the *y*-component it is a delta function.